\documentstyle[12pt]{article}
\def\hybrid{\topmargin 0pt      \oddsidemargin 0pt
	\headheight 0pt \headsep 0pt
	\textheight 9in         
	\textwidth 6.25in       
	\marginparwidth .875in
	\parskip 5pt plus 1pt   \jot = 1.5ex}

\catcode`\@=11
\def\marginnote#1{}
\newcount\hour
\newcount\minute
\newtoks\amorpm
\hour=\time\divide\hour by60
\minute=\time{\multiply\hour by60 \global\advance\minute by-\hour}
\edef\standardtime{{\ifnum\hour<12 \global\amorpm={am}%
	\else\global\amorpm={pm}\advance\hour by-12 \fi
	\ifnum\hour=0 \hour=12 \fi
	\number\hour:\ifnum\minute<10 0\fi\number\minute\the\amorpm}}
\edef\militarytime{\number\hour:\ifnum\minute<10 0\fi\number\minute}

\def\draftlabel#1{{\@bsphack\if@filesw {\let\thepage\relax
   \xdef\@gtempa{\write\@auxout{\string
      \newlabel{#1}{{\@currentlabel}{\thepage}}}}}\@gtempa
   \if@nobreak \ifvmode\nobreak\fi\fi\fi\@esphack}
	\gdef\@eqnlabel{#1}}
\def\@eqnlabel{}
\def\@vacuum{}
\def\draftmarginnote#1{\marginpar{\raggedright\scriptsize\tt#1}}

\def\draft{\oddsidemargin -.5truein
	\def\@oddfoot{\sl preliminary draft \hfil
	\rm\thepage\hfil\sl\today\quad\militarytime}
	\let\@evenfoot\@oddfoot \overfullrule 3pt
	\let\label=\draftlabel
	\let\marginnote=\draftmarginnote
   \def\@eqnnum{(\theequation)\rlap{\kern\marginparsep\tt\@eqnlabel}%
\global\let\@eqnlabel\@vacuum}  }

\def\numberbysection{\@addtoreset{equation}{section}
	\def\theequation{\thesection.\arabic{equation}}}

\def\underline#1{\relax\ifmmode\@@underline#1\else
	$\@@underline{\hbox{#1}}$\relax\fi}

\def\titlepage{\@restonecolfalse\if@twocolumn\@restonecoltrue\onecolumn
     \else \newpage \fi \thispagestyle{empty}\c@page\z@
	\def\thefootnote{\fnsymbol{footnote}} }

\def\endtitlepage{\if@restonecol\twocolumn \else  \fi
	\def\thefootnote{\arabic{footnote}}
	\setcounter{footnote}{0}}  
\catcode`@=12
\relax
\def\ie{\hbox{\it i.e. }}

\def\beq{\begin{equation}}
\def\eeq{\end{equation}}
\def\bea{\begin{eqnarray}}
\def\eea{\end{eqnarray}}

\relax
\hybrid
\begin{document}
\begin{titlepage}
\begin{center}
July~1995 \hfill    PAR--LPTHE 95/41 \\
\hfill cond-mat/9507025\\[.5in]
{\large\bf Numerical Results For The 2D Random Bond 3-state Potts
Model}\\[.5in]
        {\bf Marco Picco} \\
	{\it LPTHE\/}\footnote{Laboratoire associ\'e No. 280 au CNRS}\\
       \it  Universit\'e Pierre et Marie Curie, PARIS VI\\
       \it Universit\'e Denis Diderot, PARIS VII\\
	Boite 126, Tour 16, 1$^{\it er}$ \'etage \\
	4 place Jussieu\\
	F-75252 Paris CEDEX 05, FRANCE\\
	picco@lpthe.jussieu.fr
\end{center}

\vskip .5in
\centerline{\bf ABSTRACT}
\begin{quotation}
We present results of a numerical simulation of the 3-state Potts model
with random bond, in two dimension. In particular, we measure the critical
exponent associated to the magnetization and the specific heat. We also
compare these exponents with recent analytical computations.

\end{quotation}
\end{titlepage}
\newpage

These last years, many studies have been devoted to the problem of the
effect of randomness on 2 dimensional statistical models. The effect of
randomness is supposed to be directly related to the critical exponent of
the specific heat, $\alpha$, according to the well known Harris criterion
\cite{harris}. If $\alpha$ is positive then the disorder will be relevant,
\ie under the effect of the disorder, the model will reach a new critical
behavior at a new critical point. Otherwise, if $\alpha$ is negative,
disorder is irrelevant, the critical behavior will not change. The Harris
criterion does not give any information on the behavior of models with
$\alpha = 0$ (marginal case), and the most studied case, the 2D Ising
model, falls in this category. Due to its apparent simplicity, this model
is the most studied example of a disordered model with many analytical
results \cite{mccoy,dots1,shalaev,shankar,ludwig2,ziegler} as well as
numerical ones \cite{adsw,talapov,selke,kim}.

In the present work, we will focus on the 3-state Potts model with
disorder. The pure 3-state Potts model having a positive $\alpha$
($\alpha=1/9$), it is expected that a new critical behavior will be
obtained. The main purpose of the present work is to study numerically this
new behavior. A certain number of analytical studies of
this model predict  new critical exponents
\cite{ludwig1,DPP1,DPP2,DDPP}. In these papers, the 3-state Potts
model is considered like a generalization of the Ising model by shifting a
regularization parameter (namely the central charge of the corresponding
conformal field theory describing the behavior of the pure model at the
critical point.) Computations are performed by using a perturbed conformal
field theory with a near marginal operator associated to the
disorder. Predictions for critical exponents of the 3-state
Potts model in presence of disorder are obtained by employing
perturbative expansion with the machinery of the
renormalisation group. In particular, the
exponent associated to the spin-spin correlation function  is given by
\beq
<\sigma(0)\sigma(R)> \simeq R^{-2\Delta^{'}_{\sigma}}
\eeq
with
\beq
\label{ndim}
2\Delta'_{\sigma} = 2\Delta_{\sigma} - {27\over16}
{\Gamma^2(-{2\over3})\Gamma^2({1\over6})\over\Gamma^2(-{1\over3})
\Gamma^2(-{1\over6})} \epsilon^3 + O(\epsilon^4),
\eeq
where $\epsilon$ is
the perturbation parameter. It measures the deviation from the Ising model,
see \cite{DPP1,DPP2} for details.)  The critical exponent associated to the
energy-energy correlation function is given by
\beq
<\varepsilon(0)\varepsilon(R)>\simeq R^{-2\Delta^{'}_{\varepsilon}} \eeq
with \beq 2\Delta'_{\varepsilon} = 2\Delta_{\varepsilon} -3\epsilon
+{9\over4}\epsilon^2 + O(\epsilon^3).
\eeq

Here, $\epsilon$ takes the value
\beq
\epsilon = -{2\over 15}
\eeq
for the 3-state Potts model. Plugging this value into the previous
expressions, we obtain
\beq
\Delta'_{\sigma}=\frac{2}{15}+0,00132+O(\epsilon ^4)
=0,13465+O(\epsilon ^4).
\eeq
It will be more convenient to express $\Delta'_{\sigma}$ in function of
$\Delta_{\sigma}=\frac{2}{15}$. Then
\beq
\label{delta}
\Delta'_{\sigma}=\Delta_{\sigma}(1+\delta)\qquad;\qquad \delta=0.0099
\simeq 0.01\; .
\eeq
For $\Delta'_\varepsilon$, we have
\beq
\Delta'_{\epsilon}=1,02+O(\epsilon ^3)
\eeq
which can be compared to the value for the pure 3-state Potts model
\beq
\Delta_{\epsilon}=0.8\; .
\eeq
It is clear at this stage that the deviation of $\Delta_{\sigma}$ due to
the disorder will be very difficult to measure, while the one of
$\Delta_\varepsilon$ should be observed easily.

$\Delta'_\sigma$ and $\Delta'_\varepsilon$ do also appear in the
magnetization and in the specific heat. Using some standard finite size
scaling arguments \cite{fss}, we have at the critical point
\bea
M(L) &\simeq& L^{-\Delta^{'}_\sigma}\\
C(L) &\simeq& L^{2-2\Delta^{'}_{\varepsilon}}.
\eea

These results were in fact obtained by using perturbative computations
around the replica symmetry solution. It can also be argued that it is
necessary to break the replica symmetry, in a fashion similar to
the Parisi solution for the mean field spin glass \cite{mpv}.
If such a solution is employed, then the modification appears only at the
third order in $\epsilon$ for the energy-energy correlation function, while
there is no modification for the spin-spin correlation
function (at least up to fourth order in $\epsilon$.) The new exponent
for the energy-energy correlation function is
given by \cite{DDPP}
\beq
\Delta''_{\varepsilon}=\Delta_{\varepsilon}-\frac{3}{2}\epsilon
+O(\epsilon^{3})=1+O(\epsilon^{3}).
\eeq

Thus we have a certain number of exponents that we can compare to the
numerical results that will be presented below.

The Hamiltonian of the simulated model is given by
\beq
H=-\sum_{\{i,j\}} J_{ij}\delta_{\sigma_i,\sigma_j}
\eeq
where the coupling constant between nearest neighbor spins takes the value
\beq
J_{ij}=pJ_0 + (1-p) J_1.
\eeq
Without any lost of
generality, we can consider the case where $p={1\over 2}$. Then the model
is self-dual and thus the critical temperature is exactly known. It
is given by the solution of the equation
\beq
{1-e^{-\beta J_0} \over 1+(q-1)e^{-\beta J_0}}  =  e^{-\beta J_1}.
\eeq
In this Letter, we will only consider the case with a strong disorder, \ie
$J_0=1, J_1={1\over 10}$. The reason for such a choice is that there exists
a cross-over length $l_J$, depending on the strength of the disorder.
It is only for distance larger than $l_J$ that we expect to measure the
critical behavior of the model with disorder. Extrapolating the results of
the 2d random bond Ising model, we expect that with the present disorder,
the cross-over length will take the value $l_J \simeq 2-5$. We will come
back later to the manifestation of this cross-over length. Thus by running
on large lattice, up to $1000^2$ for results presented here, we are quite
sure to measure the critical behavior of the model with disorder.

Monte Carlo data were obtained by using the well known Wolff cluster
algorithm \cite{wolf} as well as the Swendsen-Wang algorithm
\cite{sw}. Details on the algorithms and the simulation parameters will be
presented in a subsequent paper. Measurements were performed on a square
lattice with helical boundary conditions. Due to the very strong disorder
that we consider, we needed to have huge statistics over the number of
configurations of disorder. Simulations were performed for lattice with
size ranging from $L=2$ to $L=1000$. For the magnetization, the number of
configurations of disorder were $100 000$ for $L=2-50$, then $40 000$ for
$L=100$ to $4 000$ for $L=1000$. For each of these configuration of
disorder, measurements were taken over $t_1=1000$ configurations. For the
specific heat, the number of configurations need to be larger, namely
$t_2=20000$.  $10000$ configurations of disorder were simulated for
$L=2-50$, then from $1000$ for $L=100$ to $100$ for $L=1000$. Finally,
statistical errors were computed by taking the mean value over the
configurations of disorder. The values of $t_1$ and $t_2$ were determined
in order that the thermal fluctuations are smaller than the one coming from
the distribution of disorder.

The first result that we present is the Log-Log plot of the magnetization
versus the lattice size $L$ at the critical point (see Fig.\ 1). This plot
exhibits a perfect scaling behavior with a very small deviation between
$L=2$ and $L=5$. It is the manifestation of the cross-over length and
justifies our previous assumption on its value. This is even more obvious
on the second figure. There we plot $M(L) L^{\Delta_{\sigma}}$ {\it vs.\
}$\ln(L)$. We see a jump in the cross-over region for $2\leq  L
\leq 5$. Outside this region we observe a rather good plateau.  We now
compute $\Delta'_\sigma$ inside the plateau region with the
parametrization of Eq. \ref{delta} :
\beq
\Delta'_\sigma = 0.1337\pm 0.0007
= \Delta_\sigma (1 + \delta)\qquad ; \qquad \delta = 0.003\pm 0.005\; .
\eeq
This is to be compared with the analytical value
\beq \delta = 0.01\; .
\eeq
At this level of precision, it is too difficult to draw any
conclusion on the validity of the computed value {\it vs.\ } the measured
one. The only obvious fact is that $\Delta'_\sigma$ is very close to
$\Delta_\sigma$. With the statistics that we present here, it is not
possible to differentiate these two exponents.

For the specific heat, the situation is different and we see
clearly the influence of disorder.  For the pure 3-state
Potts model
\beq
C(L) \simeq L^{2-2\Delta_\varepsilon} \simeq L^{0.4},
\eeq
while for the model with disorder we obtain a negative exponent (see
Fig.\ 3). The only possible parametrization of the curve is the one
with a cusp, \ie
\beq
C(L) = A + B L^{-x}.
\eeq
With such a parametrization,
we obtain
\beq
x= 0.13 \pm 0.04
\eeq
Using the relation
$x=2\Delta'_\varepsilon - 2 $ gives
\beq
\Delta'_\varepsilon = 1.065 \pm 0.02\; .
\eeq
This result should be compared with the analytical values
\beq
\Delta'_\varepsilon = 1.02 + O(\epsilon ^3)\qquad\hbox{and}\qquad
\Delta''_\varepsilon = 1 + O(\epsilon ^3).
\eeq

In conclusion, we have measured the critical exponent of the magnetization
of the 3 state Potts model in the presence of disorder. We have shown that
its value does not differ significantly from the pure case value.
Comparisons with analytical computations of this exponent \cite{DPP1,DPP2}
are beyond the level of precision of the present simulation.  For the
critical exponent associated to the specific heat (or the energy-energy
correlation function), the numerical value obtained in Eq. is very close to
the analytical one \cite{ludwig1,DPP1,DPP2,DDPP}. The result given by the
numerical simulation seems to agree better with the analytical value
obtained with a replica symmetry ansatz ($\Delta'_\varepsilon = 1.02 +
O(\epsilon ^3)$) than the one given by the replica symmetry breaking ansatz
$\Delta''_\varepsilon = 1 + O(\epsilon ^3)$.

\noindent{\large\bf Acknowledgements}

I am grateful to Vl.~Dotsenko and P.~Pujol for useful discussions.

\noindent{\large\bf Note}

As this work was being completed I became aware of a related work by
J.-K.~Kim which deals with a similar problem. In particular, Kim also
measured the exponent of the magnetisation (the parameter $\eta$ of the
magnetic susceptibility in fact) for the case with weak disorder and finds
it unchanged from the case without disorder.

\newpage
\topmargin -.3in
\flushbottom
\begin {flushleft}
{\Large \bf Figure Captation}
\end{flushleft}
\begin{itemize}
\item[Fig.~1] $\ln Mag(L)$ {\it vs.\ } $\ln L$ for 3-state Potts model with
disorder 1/10
\item[Fig.~2] $Mag(L)\times L^{2/15}$ {\it vs.\ } $\ln L$ for 3-state Potts
model with disorder 1/10
\item[Fig.~3] $C(L)$ {\it vs.\ } $L$ for 3-state Potts model with
disorder 1/10, with a best fit.
\end{itemize}
\end{document}